\documentclass[prl,aps,twocolumn,superscriptaddress,showpacs]{revtex4}
\usepackage{amsbsy}
\usepackage{amssymb}
\usepackage[dvips]{graphicx}

\begin{document}
\title{Critical thermodynamics of the two-dimensional $\pm J$ Ising spin glass}

\author{J. Lukic}
\affiliation{Dipartimento di Fisica, SMC and UdR1 of INFM, INFN, 
Universit\`a di  Roma {\em La Sapienza}, P.le Aldo Moro 2, 00185 Roma, Italy.}

\author{A. Galluccio}
\affiliation{CNR IASI,
Viale Manzoni 30, 00185 Roma, Italy.}

\author{E. Marinari}
\affiliation{Dipartimento di Fisica, SMC and UdR1 of INFM, INFN, 
Universit\`a di  Roma {\em La Sapienza}, P.le Aldo Moro 2, 00185 Roma, Italy.}

\author{O. C. Martin}
\affiliation{Laboratoire de Physique Th\'eorique et Mod\`eles Statistiques,
b\^atiment 100, Universit\'e Paris-Sud, F--91405 Orsay, France.}

\author{G. Rinaldi}
\affiliation{CNR IASI,
Viale Manzoni 30, 00185 Roma, Italy.}

\date{\today}

\begin{abstract}
We compute the exact partition function of $2d$ Ising spin glasses
with binary couplings.  In these systems, the ground state is highly
degenerate and is separated from the first excited state by a gap of
size $4J$. Nevertheless, we find that the low temperature specific
heat density scales as $\exp (- 2 J / T )$, corresponding to an
``effective'' gap of size $2J$; in addition, an associated cross-over
length scale grows as $\exp ( J / T )$. We justify these scalings via
the degeneracy of the low lying excitations and by the way low energy
domain walls proliferate in this model.
\end{abstract}

\maketitle

Spin glasses~\cite{MezardParisi87b,Young98} are strongly frustrated
materials that have challenged statistical physicists since many
years.  In particular, there is still no consensus on the nature of
these materials' phase diagram, a very basic issue. Surprisingly, open
questions remain even in the case of two-dimensional spin glasses. For
instance, there is a long-standing
dispute~\cite{WangSwendsen88,SaulKardar93,SaulKardar94} concerning the
$\pm J$ Ising spin glass: it is not clear what kind of singularity
arises in its free energy at the critical temperature.

In this work we reconsider the nature of these singularities using
recently developed methods~\cite{GalluccioLoebl00,GalluccioLoebl01}
for computing the exact partition function of square lattices with
periodic boundary conditions, focusing on the low $T$ scaling
properties of the model with binary couplings. We show that although
the energy ``quantum'' of excitation above the ground state is $4J$,
such excitations behave as {\em composite} particles; in fact the
specific heat near the critical point scales as if the elementary
excitations were of energy $2J$. We justify this picture using
properties of excitations and domain walls in this model. Finally, the joint
temperature and size dependence shows the presence of a characteristic
temperature-dependent length that grows as $\exp(J/T)$, in agreement
with hyperscaling.

\paragraph*{The model and our measurements ---} The 
Hamiltonian of our two-dimensional ($2d$) spin glass is
\begin{equation}
\label{eq:H}
H_{J}(\{\sigma_{i}\}) \equiv
  -\sum_{\langle ij \rangle} J_{ij}\sigma_{i}\sigma_{j}
\end{equation}
where the sum runs over all nearest neighbor pairs of Ising spins
($\sigma_{i}=\pm 1$) on a square lattice of volume $V = L \times L$
with periodic boundary conditions.  The quenched random couplings
$J_{ij}$ take the value $\pm J$ with probability $1/2$.  The partition
function at inverse temperature $\beta\equiv T^{-1}$ is
$Z_{J}=\sum_{\{\sigma_{i}\}}e^{-\beta H_{J}(\{\sigma_{i}\})}$ and can
be written as
\begin{equation}
\label{eq:Zbeta}
Z_J(\beta)=e^{2 L^2 \beta J} \ P_J(X=e^{-2\beta J})\ .
\end{equation}
Here $P_J(X)$ is the polynomial whose coefficient of $X^p$ is the
number of spin configurations of energy $E = (-2 L^2 + 2 p) J$.
Saul and Kardar~\cite{SaulKardar93,SaulKardar94} showed that
determining $P_J$ can be reduced to computing determinants which they
did using exact arithmetic of arbitrarily large integers. More
recently a more powerful approach has been
developed~\cite{GalluccioLoebl00,GalluccioLoebl01}, based on the use
of {\em modular arithmetic} to compute pfaffians. With this algorithm,
one first finds the coefficients modulo a prime number, thereby
avoiding costly arbitrary precision arithmetic. Then the computation
is repeated for enough different primes to allow the reconstruction of
the actual (huge) integer coefficients using the Chinese remainder
theorem.

The algorithm proposed and implemented in
\cite{GalluccioLoebl00,GalluccioLoebl01} is powerful enough to solve
samples with $L \approx 100$; the total CPU time needed to compute
$Z_J$ grows approximately as $L^{5.5}$. In our study we have
determined $Z_J$ for a large number of disorder samples at different
lattice sizes: for instance we have $400000$ samples at $L=6$,
$100000$ at $L=10$, $10000$ at $L=30$, $1000$ at $L=40$ and $300$ at
$L=50$. The total computation time used is equivalent to about 40
years of a 1.2 GHz Pentium processor.  For each sample we derive from
$Z_J$ various thermodynamic quantities such as the free energy
$F_J(\beta) = - \beta^{-1} \ln Z_J$, the internal energy $U_J(\beta) =
\langle H_J \rangle$, and the specific heat $dU_J/dT$.  We also study
in detail the number of ground states and of excited states.  Note
that flipping any spin changes the energy by $0$, $\pm 4J$ or $\pm
8J$; the gap between the ground state and the first excited state is
thus $4J$.

\paragraph*{Low temperature behavior of $c_V$ ---}
The study of $2d$ Ising spin glasses has a long history. We will only
discuss here results about the $\pm J$ model.  It is generally agreed
that this model is paramagnetic for $T>0$, spin-glass ordering arising
only as $T \to 0$.  The critical region thus corresponds to $T \to
T_c=0$.  Since there is an energy gap $4J$, the free energy should
have a singularity of the form $\exp(-4J/T)$.  This is difficult to
check, in particular via Monte Carlo where the free energy is not
directly measurable. Instead, it is better to concentrate on the
specific heat density $c_V$. For that observable, the difference
between the models with bimodal ($J_{ij}=\pm J$) and continuous
couplings is striking: in the first case $c_V$ goes to zero rapidly as
$T\to 0$ while in the second there is a clear linear behavior.

Even though our computations provide us with the free energy, we also
prefer to work with $c_V$.  Note that $c_V$ is related to a second
derivative of the free energy so the corresponding singularities are
directly related.  Also, $c_V$ should provide a cleaner signal as
irrelevant ``constants'' such as the ground state energy that
fluctuate from sample to sample have been subtracted out. Consider now
any given sample.  As $T \to 0$, we have the scaling
\begin{equation}
\label{eq:cv_scaling_naive}
c_V \equiv \frac{\langle \  \left[ H - \langle H \rangle \right]^2 \ \rangle }{L^2 T^2} \thickapprox
\frac{16 J^2 \ e^{S_1 - S_0} \ e^{- 4 J / T} }{L^2 T^2} 
\end{equation}
where $S_0$ and $S_1$ are the logarithms of the degeneracies of the
ground state and first excited state energy levels for the given
sample. ($S_0$ and $S_1$ are microcanonical entropies; we have dropped
the index $J$ denoting a sample dependence.) Note that $4J$ appears
because it is the energy gap in our system.  It thus seems unavoidable
that $c_V$ will have an $\exp(-4 J / T)$ singularity. Surprisingly,
in 1988, Wang and Swendsen~\cite{WangSwendsen88} postulated that
instead
\begin{equation}
\label{eq:cv_scaling}
c_V \thickapprox T^{-p} \ \exp\left(-A J/T\right)
\end{equation}
with $A=2$. They performed a Monte Carlo study in which
$A \approx 3$ for most of the temperatures they
could access, but their effective $A$
drifted and their final prediction was $A=2$ from an analogy with 
a one dimensional model (we shall come
back to this later). This issue was taken up a few years later by
Saul and Kardar~\cite{SaulKardar93,SaulKardar94} who
claimed $A=4$; their work
is based on exact computations of partition functions
and thus does not suffer at low $T$ from the
thermalization problems of the Monte Carlo approach.
We are aware
of no specific heat measurements in this model since.
How could $A$ {\em not} be $4$? The subtlety is that we
must take $L \to \infty$ at fixed $T$, and only \emph{after} can we
take $T \to 0$; indeed Eq.~(\ref{eq:cv_scaling}) assumes $L = \infty$
whereas Eq.~(\ref{eq:cv_scaling_naive}) assumes
$T\to 0$ at fixed $L$.

\begin{figure}
\includegraphics[width=5cm, height=8cm,angle=270]{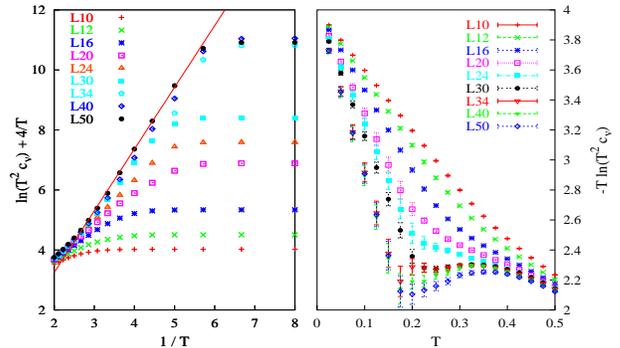}
\caption{On the left:
$\ln\left(T^2c_V\right)+4/T$ versus $1/T$.
On the right:
$-T\ln\left(T^2c_V\right)$ versus $T$.
\protect\label{fig:CVAB}} 
\end{figure}

Using the algorithm in \cite{GalluccioLoebl00,GalluccioLoebl01},
together with the availability of cheap and powerful computers,
we have extended significantly the study of Saul and Kardar.
For the sake of comparison, they had $80$ samples at $L=20$, 22, and 24, 
and 4 samples at $L=32$ and $L=36$. (They also had
samples for $L \le 18$.) We
go much beyond that, both in lattice sizes and in the number
of samples we consider. 
In the left part of
Fig.~\ref{fig:CVAB} we show our first analysis of $c_V$ as follows.
Let's set $J=1$. When $T\to 0$, if na\"{\i}ve scaling ($A=4$) holds,
$\ln\left(T^2c_V\right)+4/T\approx\mbox{const}$, while
$\ln\left(T^2c_V\right)+4/T\sim\frac{4-A}T$ if $A\ne
A_{\mbox{na\"{\i}ve}}$ and $p= p_{\mbox{na\"{\i}ve}}=2$. 
(The $c_V$ resulting from our exact
partition function computations has been averaged over
disorder samples.) In the plot we see that 
for any given lattice size, when $T$ becomes small enough 
there is a saturation toward
the na\"{\i}ve scaling behavior, i.e., the points go to a constant
value. The physically relevant regime is the thermodynamic
limit, given by the envelope of these curves; this envelope
does appear and seems to be linear
in $1/T$. Note
that the envelope emerges only on quite large lattices ($L\ge 30$); 
because of this, the true scaling escaped detection by 
Saul and Kardar. 
The straight line in the left part of
Fig.~\ref{fig:CVAB}
is our best linear fit to the $L=50$ data when
$\beta \in \left[2.5,5.5\right]$. It is a very satisfying fit and gives 
$A = 2.02\pm 0.03$, close to the integer value $A=2$.

We can also present the data in a slightly different fashion.
In the right part of
Fig.~\ref{fig:CVAB} we plot $-T\ln\left(T^2c_V\right)$ versus $T$. Here the
coefficient $A$ is given by the intercept of the envelope's
extrapolation to $T=0$, the left axis of the picture. We can
distinguish three regions. The first region is for very low $T$ values. Here
the na\"{\i}ve (non-thermodynamic scaling) with $A=A_{\mbox{na\"{\i}ve}}=4$ is very
clear. This region, where the intercept at $T=0$ is $4$, shrinks to
zero with increasing lattice size.  In a second region we have the
physical scaling; for the large lattice sizes we have, the value
$A\approx2$ emerges. Notice that this is the same region where in
the left part of 
Fig.~\ref{fig:CVAB} the $L=50$ data lie on a straight line.
The third and last region corresponds to ``high'' $T$
($T\gtrsim 0.4$) where one is far from the critical point and 
no scaling is apparent.

Our conclusion here is that thanks to the larger sizes
available to us and to a technique that does not suffer from
low temperature critical slowing down, the thermodynamic
scaling of $c_V$ is now finally clarified.

\paragraph*{Ground state properties ---}
Our computations also give the
ground state energies and degeneracies. Theoretical 
arguments~\cite{BouchaudKrzakala03} suggest that the
mean ground state energy density $e_0$ has power
corrections in $1/L$:
\begin{equation}
\label{eq:Theta}
e_0(L) = e_0^* + a \  L^{-2+{\Theta}^{(e)}}\ \ .
\end{equation}
We have $e_0^*=-1.4017(3)$ 
which agrees well with previous work.
We also find $\Theta^{(e)}=-0.08(7)$; note that the prediction 
in \cite{BouchaudKrzakala03} is that 
$\Theta^{(e)}=\theta_{DW}$, the exponent associated
with domain wall energies. Following the work of Hartmann and
Young~\cite{HartmannYoung01}, there is general agreement that
$\theta_{DW}=0$ in the $2d$ $\pm J$ model. Thus
our estimate for $\Theta^{(e)}$ is in excellent agreement with the 
conjecture in~\cite{BouchaudKrzakala03}.

We have performed a similar study for the mean ground state entropy
density $s_0(L)$.  We find $s_0^*=0.0714(2)$ which compares well with
the recent work of~\cite{BlackmanGoncalves98} in which $s_0^*=
0.0709(4)$. The fit also gives $\Theta^{(s)}=0.42(2)$, though if we
take into account systematic effects we cannot rule out
$\Theta^{(s)}=1/2$.  We believe that this large value, unrelated to
$\theta_{DW}$, denotes the presence of a subtle organization of the
ground states.

\paragraph*{Anomalous density of excitations ---}
The microcanonical entropy $S(E)$ of an energy level $E$
is defined as the logarithm of the number of spin configurations
having exactly that energy. Clearly, $S(E)$ is obtained from
the knowledge of $P_J$ as computed in Eq.~(\ref{eq:Zbeta}).
Of major interest is $S_1-S_0$,
the increase of entropy when going from the ground state energy
$E_0$ to the lowest excitation energy.
In the pure ferromagnetic model, the lowest 
excitation corresponds to taking the ground
state (all spins parallel) and flipping a single
spin. This gives $S_1 - S_0 = \ln\left(V\right)$. One says that
the excitations are ``elementary'', and the system at low temperature is
accurately described as a gas of independent excitations.

\begin{figure}
\includegraphics[width=5cm, height=8cm,angle=270]{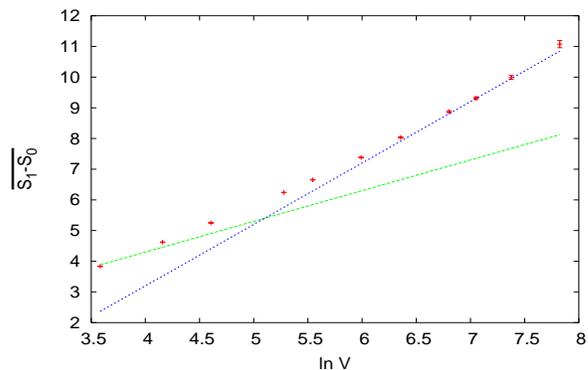}
\caption{$\overline{S(E_{0}+4J)-S(E_{0})}$ versus $\ln\left(V\right)$ and
the functions $\ln\left(V\right)$ and $2\ln\left(V\right)$.}
\protect\label{fig:DS}
\end{figure}

The situation changes dramatically when a large enough
fraction of $J_{ij}$ are negative, taking the system from a
ferromagnetic to a spin-glass phase. In such a phase,
the large $V$ law of $S_1 - S_0$ is modified.
This is illustrated in Fig.~\ref{fig:DS} where we
plot our numerical estimate of $\overline{S_1 - S_0}$ as
a function of $\ln\left(V\right)$. The
dotted line is $\ln\left(V\right)$ while the dash-dotted one is
$2 \ln\left(V\right)$. We see
that the true scaling behavior emerges only
for large lattices and that the large $V$ behavior is 
compatible with a $2 \ln\left(V\right)$ growth.
How can one interpret this anomalous growth?
Imagine classifying all excitations in terms of the
size of the cluster of spins flipped when comparing to 
a given ground state configuration. (Naturally, one may
ignore all clusters of zero excitation energy, and 
it is enough to focus on connected clusters.)
Just as in the ferromagnetic case, 
some of the excitations correspond to single spin flips;
there are $O(V)$ such objects. 
For any bounded-size cluster, the number of objects
is $O(V)$, leading to $S_1 - S_0 \approx \ln\left(V\right)$.
Since one has instead $\overline{S_1 - S_0} \approx 2\ln\left(V\right)$, 
finite-size clusters are
irrelevant: necessarily large scale excitations {\em dominate}
the set of excitations of lowest energy.

It is important to understand the nature
of these large scale excitations, but unfortunately our 
computational approach does not give us configurations, it
merely counts their number. 
There are other 
ways to gain insight into this problem. To begin, we consider
as in~\cite{WangSwendsen88} 
an analogy with the $1d$ pure Ising model.
In that system, when using periodic boundary conditions, the lowest
excitation is composite, corresponding to a pair of kinks with a total
energy $4J$; however the ``true'' elementary excitations are {\em
single} kinks, necessarily absent when using periodic boundary
conditions.  It is easy to see that for this $1d$ model the quantity
$S_1-S_0$ grows as $2 \ln\left(V\right)$, {\em i.e.}, as in our $2d$
system. 

How may objects of energy $2J$ appear in our 2d lattices with periodic
boundary conditions for which the gap is $4J$?  To answer this
question, consider in a ground state configuration any connected
cluster of spins and associate to its surface the corresponding closed
path $\cal P$ on the dual lattice~\cite{Schuster79}. 
(The cluster's surface is the set
of edges connecting the cluster to its complement.) When flipping the
cluster, the change in the configuration's energy comes only from
those bonds crossing $\cal P$; in fact, for each such bond that is
satisfied ($J_{ij}S_i S_j = 1$), the energy increases by $2J$, and
otherwise it is decreased by $2J$. It is easy to see that all clusters
lead to $\cal P$ with an even number of bonds, and thus excitation
energies are quantized in units of $4J$. However there are closed
paths that are \emph{not} associated with clusters: an example is a
path that winds around one of the directions of the lattice!  Such
topologically non-trivial paths are called domain walls; when
comparing periodic and anti-periodic boundary conditions, the set of
bonds in the ground state that are changed from satisfied to
unsatisfied or vice-versa form exactly such a path. 
When $L$ is odd,
domain walls have energies $\pm 2J$, $\pm 6J$, \ldots and the
the quantum $2J$ appears.  Of course, to have a physical excitation,
one needs to introduce domain walls in \emph{pairs}; then the flipped
cluster of spins is topologically a strip with a surface in two
pieces, one for each domain wall, while its energy is a multiple of
$4J$. Note that this is exactly what happened in the one dimensional
case, the domain walls there being simply kinks which also arise in
pairs.

To justify the anomalous scaling of $S_1 - S_0$, we appeal to the fact
realized a few years ago by Hartmann and Young~\cite{HartmannYoung01}
that low energy domain walls proliferate in the $\pm J$ spin glass.
Let $\delta S$ be the typical entropy of a single domain wall; if we
focus on those excitations of energy $4J$ associated with two domain
walls of energy $2J$, we have an excess entropy $\Delta S = 2 \delta
S$. Our data thus suggest $\delta S = \ln\left(L^2\right)$; this law
can be interpreted by saying that $\delta S$ is the sum of a
$\ln\left(L\right)$ term coming from the $L$ possible mean transverse
positions of the domain wall and of an additional $\ln\left(L\right)$
term coming from the degeneracy (proliferation) at a given position.
To extend this reasoning to the case of $L$ even, we first
remark that the domain walls there have energies $0$, $4J$, \ldots To
have a ``strip'' excitation, we need one domain wall of energy $4J$
and one of $0$ energy.  Undoubtedly, the entropy of these domain walls
increases with their energy; a simple pattern is obtained if we
conjecture that the excess entropy increases by $\ln\left(L\right)$
every time the energy increases by the quantum $2J$.  If this is so,
the first domain wall will contribute $3\ln\left(L\right)$ to the
excess entropy and the second $\ln\left(L\right)$, leading again to
the desired $2 \ln\left(V\right)$ result. Such a conjecture is quite
elegant and should be amenable to testing using a recent Monte Carlo
method~\cite{Houdayer01}.

\paragraph*{Finite-size scaling ---}
Given the result for $S_1-S_0$, we go back 
to Eq.~(\ref{eq:cv_scaling_naive}) to understand the 
finite-size scaling of $c_V$. When $T\to 0$ and $L \to \infty$
simultaneously, standard arguments lead to
\begin{equation}
\label{eq:cv_fss}
T^2 c_V(L,T) \ e^{2 \beta J} \thickapprox 
{\cal F}\left[ L/\Lambda(T) \right] \ .
\end{equation}
\begin{figure}[b]
\includegraphics[width=5cm, height=8cm,angle=270]{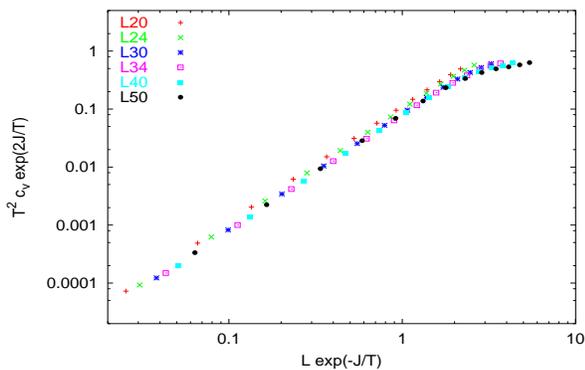}
\caption{Data collapse plot of the finite size scaling function
${\cal F}\left[ L/\Lambda(T) \right]$ with
$\Lambda(T) = exp(J/T)$ and $J=1$.}
\protect\label{fig:FSS}
\end{figure}
Here $\Lambda(T)$ is a temperature dependent length
that determines the cross-over between the 
thermodynamic scaling of $c_V$ (going as $\exp\left[-2 \beta J\right]$
when $L=\infty$) and the
``na\"{\i}ve'' scaling as in Eq.~(\ref{eq:cv_scaling_naive}).
$\cal F$ is a finite-size scaling function; when its argument
is large, $L \gg \Lambda(T)$, we recover the thermodynamic limit and thus
necessarily $\cal F$ must tend toward a constant. (Since
$c_V$ is intensive, the $L$ dependence must drop out.)
On the contrary, when $L \ll \Lambda(T)$, 
we recover the behavior
of Eq.~(\ref{eq:cv_scaling_naive}) where $c_V$ goes as
$\exp(-4 \beta J)$ \emph{but} diverges as $L \to \infty$. Interestingly,
in this regime $c_V$ is not self-averaging and so one should
apply finite-size scaling for the whole probability distribution
of $c_V$. 
Just as before where we computed
$\overline{S_1-S_0}$ and not $\log(\overline{\exp{(S_1-S_0)}})$:,
we focus on the typical behavior and so we consider
the median rather than the average of $c_V$ (this distinction is relevant only
in the very low $T$, unphysical region, while it is irrelevant in our
scaling region for $T$, say, close to $0.3$).  When using this data we
have a very reasonable data collapse, consistent with
Eq.~(\ref{eq:cv_fss}) as shown in Fig.~\ref{fig:FSS}.
We find that this median scales as
$L^2 \exp(-2 \beta J)$ at low $T$ and thus
${\cal F}(x) \approx x^2$ as $x \to 0$. 
This then gives
\begin{equation}
\Lambda(T) \sim \exp(\beta J) \ .
\label{eq:Lambda}
\end{equation}

\paragraph*{Summary and discussion ---}
We have investigated the critical thermodynamics
of the $2d$ Ising spin glass with binary couplings. 
Our main conclusion is that the specific heat density
scales as $c_V\sim\exp(-2J\beta)$.
This scaling is 
``anomalous'' in the sense that
it does not follow from the size of the energy gap
(which is $4J$). To find this scaling law, it is necessary
to go to rather large systems, $L \ge 30$.
We also found that the typical
degeneracy of the first excited level grows
about $L^4$ times faster than that of the ground state
level. We believe this high degeneracy
has its roots in the proliferation of domain walls,
two domain walls enabling one to define a composite excitation. Such
a picture justifies the analogy with kink pairs proposed many 
years ago by Wang and Swendsen~\cite{WangSwendsen88}:
each domain wall may indeed play the role of a kink,
albeit with an additional entropy contribution.
Finally, using finite-size scaling, we
found a cross-over length scale $\Lambda(T)$
that grows as $\exp(J/T)$; this divergence is exactly as
expected from hyperscaling arguments.

This work was supported in part by the European Community's
Human Potential Programme under contracts 
HPRN-CT-2002-00307 (DYGLAGEMEM) and
HPRN-CT-2002-00319 (STIPCO).

\bibliographystyle{apsrev}

\bibliography{/home/martino/Papers/Bib/references}

\end{document}